\documentclass[preprint,12pt]{elsarticle}

\usepackage{amsmath,amssymb}
\usepackage{graphicx}
\usepackage{booktabs}
\usepackage{placeins}
\usepackage[colorlinks=true,linkcolor=blue!50!black,citecolor=blue!50!black,urlcolor=blue!50!black]{hyperref}
\usepackage{xcolor}

\newcommand{\nub}{\bar\nu}
\newcommand{\dd}{\mathrm{d}}
\newcommand{\bOmega}{\boldsymbol{\Omega}}

\journal{Journal of High Energy Astrophysics}

\begin{document}

\begin{frontmatter}

\title{Neutrino pair annihilation in curved spacetime: a fast moment method and the geometry
of flavour conversion}

\author[udesc]{Rafael C. R. de Lima\corref{cor}}
\ead{rafael.lima@udesc.br}
\cortext[cor]{Corresponding author}
\author[udesc]{Luciano de Oliveira Junior}
\ead{luciano.junior1999@edu.udesc.br}
\author[udesc]{Pedro Martins Bailer}
\ead{pedromartinsbailer@gmail.com}
\address[udesc]{Departamento de Física, Universidade do Estado de Santa Catarina (UDESC),
Joinville, SC, Brazil}

\begin{abstract}
Neutrino--antineutrino annihilation above black-hole accretion discs is shaped both by
general-relativistic transport and by neutrino flavour conversion. Whether these effects can be
treated independently depends on how gravity weights the different flavour populations. We
investigate this interplay using backward ray tracing in Schwarzschild and Kerr spacetimes,
combined with an exact angular-moment formulation of the annihilation kernel that we validate
against the analytic spherical solution and published disc calculations. We first establish that
flavour-conversion corrections are independent of the spacetime geometry when conversion is
uniform and the initial heavy-lepton neutrino emission is negligible. For uniform conversion, this
separability remains approximately valid when the flavour-dependent emitting regions overlap:
compared with the same discs in flat space, moving with the same local velocities, gravity
enhances the deposited energy by factors of 1.4--5, while changing the relative conversion effect
by no more than 6\% for most temperature profiles considered; conversion that varies across the
disc raises this to 22\% around a rapidly spinning hole. A qualitatively different regime emerges
when heavy-lepton neutrinos originate from an inner ring separated from the electron-flavour
emitting regions. Relativistic bending and energy shifts preferentially enhance annihilations
involving inner-ring neutrinos, producing relativistic conversion factors up to 2.4 times the
flat-space ones. In one configuration, flavour equipartition reduces the deposition by 54\% in
flat space but increases it by 11\% in Kerr spacetime. These results identify the spatial arrangement of flavour-dependent
neutrinospheres as a key determinant, in the models considered, of whether flavour conversion and
relativistic transport can be treated separately.
\end{abstract}


\end{frontmatter}

\section{Introduction}\label{sec:intro}

Copious neutrino emission accompanies the birth of neutron stars in core-collapse supernovae
and the hyper-accreting black holes and massive neutron stars formed in compact-object
mergers and collapsars. Above the neutrinosphere, a small fraction of the neutrinos and
antineutrinos annihilate, $\nu\nub\to e^+e^-$, and deposit energy into a pair plasma. Goodman,
Dar and Nussinov showed that in a type~II supernova this process can deposit more than
$10^{51}$\,erg above the neutrinosphere~\cite{Goodman:1986we}. Neutrino annihilation has since
been considered as a driver of neutrino-heated winds and as a possible engine of the
relativistic jets of gamma-ray bursts
(GRBs)~\cite{Popham:1998ab,Janka:1995cq,Ruffert:1996by,Liu:2017kga}.

Whether annihilation can power GRBs has been debated since the 1990s. Newtonian merger
simulations found rates far below what is needed without strong
collimation~\cite{Janka:1995cq,Ruffert:1996by}, models of neutrino-cooled discs showed a steep
dependence on accretion rate and spin~\cite{Popham:1998ab,Chen:2006rra,Zalamea:2010ax}, long
events strain the energetics~\cite{Leng:2014dfa}, and the most recent general-relativistic
radiation-hydrodynamics simulations find outflows compatible with low-luminosity short GRBs and
precursors~\cite{Kawaguchi:2024naa,Kawaguchi:2025con}. Two physical ingredients shape the deposition
in these systems.

The first is general relativity (GR). Light bending enlarges the apparent size of the emitter and
favours head-on collisions, while gravitational and Doppler shifts change the neutrino energies,
to which the rate is very sensitive. Around a spherical neutrinosphere the deposition grows by
factors of 2 to 30 as the radius approaches the photon sphere~\cite{Salmonson:1999es}; above discs
and tori, global changes range from tens of per cent to a factor of 2, with larger local
effects~\cite{Asano:2000ib,Asano:2000dq,Birkl:2006mu,Harikae:2009yt,Harikae:2010yt,Perego:2017fho},
and rotation of the emitter reduces the rate~\cite{Prasanna:2001ie,Mallick:2009nvq,Kovacs:2009dv}.
The second is flavour conversion. Simulations of post-merger discs find conditions for fast
flavour instabilities, with conversions that can be
widespread~\cite{Li:2021vqj,Just:2022flt,Mukhopadhyay:2024zzl} and depend on direction and
position~\cite{Lund:2025jjo}. Since the annihilation coupling of $\nu_e\nub_e$ pairs is about 4.7
times that of heavy-lepton pairs, redistributing neutrinos among flavours changes the deposition;
flavour equipartition was estimated to reduce it by up to a factor of $\sim5$ in a hyper-accreting
disc~\cite{Uribe:2019cpq}, and vacuum oscillations above a neutrino-dominated accretion flow by up
to 19\%~\cite{Deesamer:2024gxv}.

These two ingredients are usually assessed separately: gravity reshapes the trajectories and
energies of the neutrinos, whereas flavour conversion redistributes the occupation of the
interacting species. Their effects, however, need not factorise. If electron and heavy-lepton
neutrinos originate from different parts of the disc, light bending and energy shifts weight
their contributions differently. The central question is therefore not only how much gravity
enhances annihilation, but when it changes the impact of flavour conversion itself. We address it
with a semi-analytic approach that isolates the transport: it does not follow the hydrodynamics of
the pair plasma or the dynamics of the conversion, which simulations such as those of
Refs.~\cite{Kawaguchi:2025con,Lund:2025jjo} address, and the conversion is parametrised.

Our contributions are three. (i) A method that evaluates the annihilation kernel exactly through
angular moments of the redshifted energy moments, without moment closure, combined with backward
ray tracing in Schwarzschild and Kerr spacetimes and validated against analytic and published
results, including the exact deposited Killing energy (Sections~\ref{sec:moments}--\ref{sec:numerics}
and~\ref{sec:valid}). (ii) An exact separability limit: for uniform conversion and negligible
heavy-lepton emission, the conversion factor does not depend on gravity or on the emission geometry
(Section~\ref{sec:flavour}). (iii) A map of the regimes in which flat-space conversion factors can or
cannot be used in curved spacetime: in the models considered, they can, approximately, when the
flavour-dependent emitting regions overlap and the conversion is uniform, and they can fail, up to
reversing the sign of the conversion effect,
when the heavy-lepton emission is segregated in an inner ring, for a physical reason that we
identify (Section~\ref{sec:flav-res}). The question that runs through the paper is thus: under
what conditions can flavour conversion be computed in flat space and applied to a relativistic
deposition?

\section{Energy deposition by neutrino pair annihilation}\label{sec:rate}

Let $f_\nu(\mathbf p,x)$ and $f_{\nub}(\mathbf p',x)$ be the phase-space densities of a
neutrino species and its antiparticle at the event $x$. In the frame of a local observer, the
energy deposited per unit volume and time by $\nu\nub\to e^+e^-$ is~\cite{Goodman:1986we,Salmonson:1999es}
\begin{gather}\label{eq:qdot-kinetic}
\dot q=\iint f_\nu f_{\nub}\,\{\sigma|\mathbf v_\nu-\mathbf v_{\nub}|\varepsilon\varepsilon'\}
\frac{\varepsilon+\varepsilon'}{\varepsilon\varepsilon'}\,\dd^3p\,\dd^3p',\\
\{\sigma|\mathbf v_\nu-\mathbf v_{\nub}|\varepsilon\varepsilon'\}=\frac{DG_F^2c}{3\pi}
(\varepsilon\varepsilon'-\mathbf p\cdot\mathbf p'c^2)^2 ,\nonumber
\end{gather}
where the bracket is Lorentz invariant, the electron mass is neglected, $G_F$ is the Fermi
constant and $D$ is the coupling of the pair, $D_e=1+4\sin^2\theta_W+8\sin^4\theta_W$ for
$\nu_e\nub_e$ and $D_x=1-4\sin^2\theta_W+8\sin^4\theta_W$ for each heavy-lepton flavour, with
$\theta_W$ the weak mixing angle. With
$\mathbf p=\varepsilon\bOmega/c$ and $\dd^3p=\varepsilon^2\dd\varepsilon\,\dd\Omega/c^3$,
\begin{equation}\label{eq:qdot-split}
\dot q=\frac{DG_F^2}{3\pi c^5}\iint\dd\Omega\,\dd\Omega'\,(1-\bOmega\cdot\bOmega')^2
\iint f_\nu f_{\nub}\,(\varepsilon+\varepsilon')\,\varepsilon^3\varepsilon'^3\,\dd\varepsilon\,\dd\varepsilon' .
\end{equation}
For isotropic Fermi--Dirac spectra with zero chemical potential and a common temperature $T$
over the solid angle subtended by the emitter, this gives the classic
result~\cite{Salmonson:1999es}
\begin{equation}\label{eq:qdot-thermal}
\dot q=\frac{7DG_F^2\pi^3\zeta(5)}{2c^5h^6}\,(kT)^9\,\Theta,\qquad
\Theta=\iint(1-\bOmega\cdot\bOmega')^2\,\dd\Omega\,\dd\Omega' .
\end{equation}

\paragraph{Statistical weight} Equation~\eqref{eq:qdot-thermal} follows the convention of
Ref.~\cite{Salmonson:1999es}, which takes, for each of $\nu$ and $\nub$,
$f=(2/h^3)(e^{\varepsilon/kT}+1)^{-1}$, i.e.\ a statistical weight $g_s=2$. A Fermi--Dirac surface with weight $g_s$ radiates the flux $g_s(7/16)\sigma T^4$,
whereas Ref.~\cite{Salmonson:1999es} writes the luminosity of one species as
$L=4\pi R^2(7/4)\sigma T^4$, with $\sigma=ac/4$, which corresponds to $g_s=4$. Neutrinos and
antineutrinos have one helicity state each, $g_s=1$. For a consistent weight the rate scales as
$g_s^2T^9$ and, at fixed luminosity ($T^4\propto L/g_s$), as $g_s^{-1/4}L^{9/4}$. The Newtonian
rate around a sphere is
\begin{equation}\label{eq:newt-norm}
\dot Q_{\rm Newt}=\kappa\,D\,L_{51}^{9/4}R_6^{-3/2}\times10^{51}\ {\rm erg\,s^{-1}},
\end{equation}
with $\kappa=6.20\times10^{-5}$ for $g_s=1$ and $\kappa=1.10\times10^{-5}$ with the conventions of
Ref.~\cite{Salmonson:1999es}, for one flavour pair with luminosity $L=10^{51}L_{51}$\,erg\,s$^{-1}$ in each species and
$R=10^6R_6$\,cm. Ratios such as the relativistic enhancement do not depend on this choice. We
use $g_s=1$ throughout, which also reproduces the absolute values of Ref.~\cite{Birkl:2006mu}
(Section~\ref{sec:results}).

\section{Exact factorisation into angular moments}\label{sec:moments}

Define the energy moments of the distribution arriving from direction $\bOmega$,
\begin{equation}
\mathcal E_n(\bOmega)=\int_0^\infty f(\varepsilon,\bOmega)\,\varepsilon^n\,\dd\varepsilon ,
\end{equation}
and, for any function $a(\bOmega)$ on the local sky, its angular moments of rank 0, 1 and 2,
\begin{equation}
a^{(0)}=\int a\,\dd\Omega,\qquad a^{(1)}_i=\int a\,\Omega_i\,\dd\Omega,\qquad
a^{(2)}_{ij}=\int a\,\Omega_i\Omega_j\,\dd\Omega .
\end{equation}
Expanding $(1-\bOmega\cdot\bOmega')^2=1-2\Omega_i\Omega'_i+\Omega_i\Omega_j\Omega'_i\Omega'_j$ in
Eq.~\eqref{eq:qdot-split} gives, exactly,
\begin{equation}\label{eq:factorised}
\dot q=\frac{DG_F^2}{3\pi c^5}[\mathcal M(\mathcal E_4,\bar{\mathcal E}_3)+\mathcal M(\mathcal E_3,\bar{\mathcal E}_4)],
\qquad
\mathcal M(a,b)\equiv a^{(0)}b^{(0)}-2\,a^{(1)}_ib^{(1)}_i+a^{(2)}_{ij}b^{(2)}_{ij},
\end{equation}
where bars refer to antineutrinos. $\mathcal M$ is symmetric and equals
$\iint(1-\bOmega\cdot\bOmega')^2a(\bOmega)b(\bOmega')\,\dd\Omega\,\dd\Omega'$. The expansion is an
elementary consequence of the polynomial form of the kernel; what matters is that it is exact, so
that no closure relation between moments is needed, unlike in moment transport
schemes~\cite{Dessart:2008zd}. Three consequences:
\begin{enumerate}
\item The cost of evaluating $\dot q$ at a point is that of computing 10 angular moments for
each of $\mathcal E_3$ and $\mathcal E_4$ (and for each of $\nu$ and $\nub$ if their
distributions differ). Since $a^{(2)}_{ii}=a^{(0)}$, nine of each set are independent. The cost
is linear in the number of sky directions, instead of quadratic (\ref{app:bench}).
\item No assumption on the spectra is needed: thermal, pinched or flavour-mixed distributions
enter only through $\mathcal E_3$ and $\mathcal E_4$, with no additional ray tracing.
\item Summing Eq.~\eqref{eq:factorised} over the three species with their own $D$ gives the total
rate.
\end{enumerate}
For a uniform emitter, $\mathcal E_n$ is constant on the cone it subtends, and $\mathcal M$
reduces to $\mathcal E_4\bar{\mathcal E}_3\Theta$, recovering Eq.~\eqref{eq:qdot-thermal}.

The momentum deposited per unit volume and time follows from Eq.~\eqref{eq:qdot-kinetic} with
$\varepsilon+\varepsilon'$ replaced by $(\varepsilon\bOmega+\varepsilon'\bOmega')/c$. With the
rank-three moments $a^{(3)}_{ijk}=\int a\,\Omega_i\Omega_j\Omega_k\,\dd\Omega$,
\begin{equation}\label{eq:pdot}
c\,\dot p_i=\frac{DG_F^2}{3\pi c^5}[\mathcal P_i(\mathcal E_4,\bar{\mathcal E}_3)+\mathcal P_i(\bar{\mathcal E}_4,\mathcal E_3)],
\qquad
\mathcal P_i(a,b)\equiv a^{(1)}_ib^{(0)}-2\,a^{(2)}_{ij}b^{(1)}_j+a^{(3)}_{ijk}b^{(2)}_{jk},
\end{equation}
where $\mathcal P_i(a,b)=\iint(1-\bOmega\cdot\bOmega')^2\,\Omega_i\,a(\bOmega)b(\bOmega')\,\dd\Omega\,\dd\Omega'$.
Energy and momentum together need 20 moments per energy moment (ranks 0 to 3), of which 16 are
independent, because $a^{(2)}_{ii}=a^{(0)}$ and $a^{(3)}_{iij}=a^{(1)}_j$; we store all 20.
Both $\dot q$ and $c\dot p_i$ have dimensions of energy per unit volume and time; they are the
time and space components $Q^{\hat\mu}$ of the deposited four-momentum density in the frame of
the local observer (\ref{app:einf}). For the sphere,
$\mathcal P_r$ with $a=b=1$ on the cone is twice the function $\Phi_p$ of
Ref.~\cite{Salmonson:1999es}; $\Phi_p/\Theta$ grows from $2/5$ at the surface to 1 far away.
The method keeps, point by point, the full angular information of each species and, through the
moments of each emitter, of each emitting region; this is what will let us separate their
contributions to the deposition.

\section{Transport in curved spacetime}\label{sec:transport}

We use $G=c=1$ in the geometric parts, with $M$ the mass of the central object, and restore the
physical constants for rates. Between the neutrinosphere and the annihilation point, neutrinos
stream freely along null geodesics; their masses are negligible at these energies. By Liouville's theorem $f$ is
constant along each ray, and the energy measured by an observer changes by the frequency
ratio
\begin{equation}
\mathsf g(\bOmega)=\frac{\varepsilon_{\rm loc}}{\varepsilon_{\rm em}}=
\frac{(p_\mu u^\mu)_{\rm loc}}{(p_\mu u^\mu)_{\rm em}},
\end{equation}
where $u^\mu$ are the four-velocities of the local observer and of the emitting matter. For
thermal emission at temperature $T_{\rm em}$, the arriving distribution is thermal with
temperature $\mathsf g\,T_{\rm em}$, and
\begin{equation}\label{eq:En-thermal}
\mathcal E_n(\bOmega)=\frac{g_s}{h^3}\,\Gamma(n+1)\,\eta_{\rm D}(n+1)\,[k\,\mathsf g(\bOmega)\,T_{\rm em}(\bOmega)]^{n+1},
\end{equation}
with $\eta_{\rm D}$ the Dirichlet eta function: $\Gamma(4)\eta_{\rm D}(4)=7\pi^4/120$ and $\Gamma(5)\eta_{\rm D}(5)=45\zeta(5)/2$.
If a ray traced backward from the local point does not end on the emitter (it escapes or
enters the horizon), $\mathcal E_n=0$ in that direction. In Schwarzschild spacetime the rate is
evaluated in the frame of the static observer; in Kerr spacetime, in that of the
zero-angular-momentum observer (ZAMO), which exists down to the horizon (the comparison with
Ref.~\cite{Birkl:2006mu} in Section~\ref{sec:valid} uses static observers, as they do). Relativistic transport
therefore changes not only the total deposition rate, but potentially the relative weights of
neutrino populations emitted from distinct disc regions, the ingredient that determines whether
flavour-conversion effects factorise.

\paragraph{Global rates} The deposition rates $\dot q$ and $c\dot p_i$ are the components
$Q^{\hat\mu}$ of the four-momentum deposited per unit proper volume and time
(\ref{app:einf}). Two volume integrals are in use. Salmonson and Wilson integrate the local energy over proper volume; in
Schwarzschild coordinates, with $A(r)=1-2M/r$,
\begin{equation}\label{eq:Eloc}
\dot E_{\rm loc}=\int\dot q\,\dd V,\qquad \dd V=\frac{r^2\sin\theta}{\sqrt{A}}\,\dd r\,\dd\theta\,\dd\phi .
\end{equation}
The Killing energy given to the pairs per unit coordinate time is the integral of $-Q_t$
(\ref{app:einf}); for static observers $-Q_t=\sqrt A\,Q^{\hat t}$ exactly, so
\begin{equation}\label{eq:Einf}
\dot E_\infty=\int(-Q_t)\sqrt{-g}\,\dd^3x=\int\sqrt{A}\,\dot q\;r^2\sin\theta\,\dd r\,\dd\theta\,\dd\phi ,
\end{equation}
with no approximation regarding the momentum of the pairs (in Kerr, $Q^{\hat\phi}$ enters through
frame dragging~\cite{Zalamea:2010ax}). The two integrals differ by a factor $A$ in the
integrand. Near a horizon a static observer sees the emitter compressed into a cone of
half-angle $\propto\sqrt A$ and blueshifted by $A^{-1/2}$, so that $\dot q\propto A^{-1/2}$:
$\dot E_\infty$ converges, while $\dot E_{\rm loc}$ diverges logarithmically and depends on where
the integration starts. Ref.~\cite{Birkl:2006mu} integrated $\sqrt{A}\,\dot q\,\dd V$, whose
integrand exceeds that of Eq.~\eqref{eq:Einf} by a factor $1/\sqrt A$, as noted in
Ref.~\cite{Zalamea:2010ax}. $\dot E_\infty$ is the energy budget given to the pairs; how much of it
reaches a distant observer depends on the subsequent evolution, since pairs created deep in the
potential or with inward momentum may be captured and the plasma redistributes energy
hydrodynamically. As in Ref.~\cite{Birkl:2006mu}, we also quote the integral over the region of
outward radial momentum deposition as a proxy for the energy available to an outflow.

\section{Emitters}\label{sec:schw}

In Schwarzschild spacetime a ray with impact parameter $b$ obeys the orbit
equation $\dd^2u/\dd\varphi^2=-u+3Mu^2$ with $u=1/r$, in the plane spanned by the radius vector and
the ray. At a point at radius $r$, a direction making an angle $\psi$ with the outward radial
direction in the static frame has $b=r\sin\psi/\sqrt{A(r)}$.

\subsection{Spherical neutrinosphere}
For a static, isothermal neutrinosphere of radius $R>3M$, the neutrinos reaching radius $r$
fill a cone around the outward radial direction with half-angle
$\psi_c$~\cite{Salmonson:1999es}:
\begin{equation}
\cos\psi_c\equiv x=\sqrt{1-(R/r)^2\,A(r)/A(R)},\qquad
\Theta(r)=\frac{2\pi^2}{3}(1-x)^4(x^2+4x+5),
\end{equation}
and $\mathsf g=\sqrt{A(R)/A(r)}$ is the same in every direction. At fixed luminosity at infinity,
$L_\infty=A(R)L(R)$. Integrating over proper volume, Salmonson and Wilson obtain the enhancement
$F(M/R)=\dot Q_{\rm GR}/\dot Q_{\rm Newt}\simeq2$, $4$ and $\sim30$ for $R/M=10$, $5$ and $3$. For
$r\gg R$ in the Newtonian limit, $\Theta\to(5\pi^2/12)(R/r)^8$. The spherical case is our first
benchmark.

\subsection{Thin Keplerian disc}
For a geometrically thin disc in the equatorial plane between $r_{\rm in}$ and $r_{\rm out}$, with
matter in circular Keplerian orbits ($\Omega_K=\sqrt{M/r^3}$, $u^t=(1-3M/r)^{-1/2}$), the frequency ratio
between a static observer at radius $r$ and the emitter at $r_{\rm em}$ is
\begin{equation}\label{eq:g-disc}
\mathsf g=\frac{1}{\sqrt{A(r)}\;u^t(r_{\rm em})\,[1-\Omega_K(r_{\rm em})\,L_z/E]},
\end{equation}
where $L_z/E$ is the conserved $z$-angular momentum per unit energy of the ray. For a static
(non-rotating) emitter, $\mathsf g=\sqrt{A(r_{\rm em})/A(r)}$. We consider isothermal discs, power
laws $T_{\rm em}\propto(r_{\rm em}/r_{\rm in})^{-s}$, as in Refs.~\cite{Asano:2000dq,Lambiase:2022ywp}, and
the profile of a radiatively efficient disc, $T_{\rm em}\propto\mathcal F^{1/4}$ with the Page--Thorne flux
$\mathcal F(r)$~\cite{Page:1974he}, which vanishes at the innermost stable circular orbit (ISCO),
of radius $r_{\rm isco}$, peaks near $1.5$--$2\,r_{\rm isco}$ and falls
as $r^{-3}$. The latter tests the sensitivity to the flux profile of a thin disc; it is not a
neutrino-dominated accretion flow computed with neutrino microphysics~\cite{Chen:2006rra,Zalamea:2010ax},
which we leave for future work. Rays may cross
the equatorial plane several times before reaching the disc; we follow the first crossings
and record their order.

\subsection{Kerr spacetime}
In Boyer--Lindquist coordinates, with $\Sigma=r^2+a^2\cos^2\theta$, $\Delta=r^2-2Mr+a^2$ (the
horizon is at its larger root, $r_+=M+\sqrt{M^2-a^2}$) and
$\mathcal A=(r^2+a^2)^2-a^2\Delta\sin^2\theta$, the ZAMO has lapse $\alpha=\sqrt{\Sigma\Delta/\mathcal A}$
and angular velocity $\omega=2Mar/\mathcal A$, and $\sqrt{g_{\phi\phi}}=\sqrt{\mathcal A/\Sigma}\,\sin\theta$.
A neutrino moving along $\bOmega$ in the ZAMO frame has
\begin{gather}
\frac{E}{\varepsilon_{\rm loc}}=\alpha+\omega\sqrt{g_{\phi\phi}}\,\Omega_{\hat\phi},\qquad
\lambda\equiv\frac{L_z}{E}=\frac{\sqrt{g_{\phi\phi}}\,\Omega_{\hat\phi}}{\alpha+\omega\sqrt{g_{\phi\phi}}\,\Omega_{\hat\phi}},\nonumber\\
\eta\equiv\frac{Q}{E^2}=\frac{p_\theta^2}{E^2}+\cos^2\theta\,(\lambda^2/\sin^2\theta-a^2),
\end{gather}
with $p_\theta/E=\sqrt\Sigma\,\Omega_{\hat\theta}/(\alpha+\omega\sqrt{g_{\phi\phi}}\,\Omega_{\hat\phi})$. In Mino time
$\sigma$ ($\dd\sigma=\dd\lambda_{\rm aff}/\Sigma$) the radial and polar motions decouple,
\begin{gather}
(\dd r/\dd\sigma)^2=R(r)\equiv(r^2+a^2-a\lambda)^2-\Delta[\eta+(\lambda-a)^2],\nonumber\\
(\dd x/\dd\sigma)^2=\Theta_x(x)\equiv\eta(1-x^2)+a^2x^2(1-x^2)-\lambda^2x^2,
\end{gather}
with $x=\cos\theta$. For a prograde Keplerian disc, $\Omega_K=M^{1/2}/(r^{3/2}+aM^{1/2})$ and
$u^t=(r^{3/2}+aM^{1/2})/[r^{3/4}(r^{3/2}-3Mr^{1/2}+2aM^{1/2})^{1/2}]$, and
\begin{equation}\label{eq:g-kerr}
\mathsf g=\frac{1}{(\alpha+\omega\sqrt{g_{\phi\phi}}\,\Omega_{\hat\phi})\,u^t(r_{\rm em})\,[1-\Omega_K(r_{\rm em})\lambda]},
\end{equation}
which reduces to Eq.~\eqref{eq:g-disc} for $a=0$. A disc of zero angular momentum moves with the
local ZAMO, $\mathsf g=\alpha(r_{\rm em})/\{(\alpha+\omega\sqrt{g_{\phi\phi}}\,\Omega_{\hat\phi})[1-\omega(r_{\rm em})\lambda]\}$.
The inner edge is placed at the ISCO. In Section~\ref{sec:flav-res}
the discs of the different species have different radial extents: overlapping in the inner disc,
or with the heavy-lepton emission confined to an inner ring.

\section{Flavour conversion}\label{sec:flavour}

We parametrise flavour conversion as a linear map applied to the distributions at emission,
or on a conversion surface, before transport. For three flavours with identical heavy-lepton
species ($x=\mu,\tau$),
\begin{equation}\label{eq:mixing}
f'_{\nu_e}=(1-P)f_{\nu_e}+Pf_{\nu_x},\qquad
f'_{\nu_x}=\tfrac P2f_{\nu_e}+(1-\tfrac P2)f_{\nu_x},
\end{equation}
with the analogous map for antineutrinos and parameter $\bar P$. The map conserves the total
number of neutrinos. $P=\bar P=0$ means no conversion and $P=\bar P=2/3$ means three-flavour
equipartition. For uniform $P$ and $\bar P$ it changes the net electron-lepton number by
$\Delta(N_{\nu_e}-N_{\nub_e})=-P(N_{\nu_e}-N_{\nu_x})+\bar P(N_{\nub_e}-N_{\nub_x})$, and that of each
heavy-lepton family by minus half of this amount. Here $N_s$ is the number of neutrinos of species
$s$ (one heavy flavour for $x$) emitted per unit Killing time: a Fermi--Dirac surface at comoving
temperature $T$ emits a number flux $\propto T^3$ per unit comoving area and proper time, and the
world volume of an equatorial disc has the invariant element $r\,\dd t\,\dd r\,\dd\phi$ in Kerr
spacetime, as in flat space, whatever its rotation, so that $N_s\propto\int T_s^3\,r\,\dd r$ over
the emitting range of the species. Conservation of the lepton number of every family is the
condition of the case ``mix1'' of Just et al.~\cite{Just:2022flt}, who impose it locally and per
energy bin. For the numbers emitted from the whole disc it holds if and only if
$P(N_{\nu_e}-N_{\nu_x})=\bar P(N_{\nub_e}-N_{\nub_x})$. We use this condition, restricted to
$\bar P\le1$, as an alternative to $\bar P=P$. Since the
map is linear, it acts directly on the energy moments $\mathcal E_n$ of Eq.~\eqref{eq:factorised}:
no additional ray tracing is needed, only the recombination of stored moments. Each converted
neutrino then streams freely from its emitter: we neglect absorption and scattering after
conversion, also where the converted neutrino crosses matter that is opaque to its new flavour,
such as a heavy-lepton neutrino of an inner ring, converted into $\nu_e$, that crosses the
$\nu_e$-emitting disc. The conversion does not change the emitting surfaces; it is phenomenological
and does not replace a solution of collective flavour evolution.

\paragraph{Conversion factor and its gravitational change} We measure the effect of conversion by
the factor $\mathcal R(P,\bar P)=\dot E_\infty(P,\bar P)/\dot E_\infty(0,0)$, the ratio of the deposited
Killing energy with and without conversion (Section~\ref{sec:transport}), and the effect of gravity on
it by
\begin{equation}\label{eq:delta}
\delta=\max_{0\le P\le2/3}\,|\mathcal R_{\rm GR}(P)/\mathcal R_{\rm flat}(P)-1| ,
\end{equation}
where $\mathcal R_{\rm flat}$ is computed for the same emitters in flat space, and $\bar P=P$, so
that $P=2/3$ is equipartition (Table~\ref{tab:flavour} also gives $\delta$ with $\bar P$ fixed by
lepton-number conservation). The emitters are identified ring by ring through their coordinate
radius $r$ (the Boyer--Lindquist radius in Kerr spacetime), with the same temperature in the frame
of the gas and the same local velocity: the flat-space disc rotates with the speed of the
relativistic Keplerian flow relative to the local ZAMO,
$v(r)=[\Omega_K(r)-\omega]\sqrt{g_{\phi\phi}}/\alpha$ at the equator, and $\Omega=v/r$. In
Schwarzschild spacetime $v=[M/(r-2M)]^{1/2}$; at the ISCO $v=0.50$, 0.62 and 0.60 for $a=0$, 0.9 and
0.99, whereas the Newtonian Keplerian speed $(M/r)^{1/2}$ is 0.41, 0.66 and 0.83. With the local
kinematics fixed, $\delta$ measures the effect of light bending, gravitational redshift and frame
dragging; we also quote it with Newtonian Keplerian speeds in flat space, to show how much the
choice of flat-space kinematics matters (Section~\ref{sec:flav-res}). The denominator is positive: in flat space
$\dot E_\infty$ is the volume integral of $\dot q\ge0$, and Eq.~\eqref{eq:mixing} keeps the
distributions non-negative for $0\le P,\bar P\le1$. If $\delta$ is small, $\dot E_\infty^{\rm GR}(P)\simeq\dot E_\infty^{\rm GR}(0)\,\mathcal R_{\rm flat}(P)$:
the conversion factor can be computed without gravity and applied to a relativistic rate computed
without conversion.

\paragraph{Common neutrinosphere} If all species share the emitting surface and its
temperature profile, up to temperature scales $T_e$, $T_{\bar e}$, $T_x$, the geometric factor is
common to all species. The ratio of deposition rates with and without conversion is then the
algebraic expression
\begin{equation}\label{eq:ratio-common}
\mathcal R(P,\bar P)=\frac{\sum_{s}D_s[\mathcal E'_{4,s}\bar{\mathcal E}'_{3,s}+\mathcal E'_{3,s}\bar{\mathcal E}'_{4,s}]}
{\sum_{s}D_s[\mathcal E_{4,s}\bar{\mathcal E}_{3,s}+\mathcal E_{3,s}\bar{\mathcal E}_{4,s}]},
\end{equation}
where $\mathcal E_{n,s}$ stands for the scale $T_s^{n+1}$ of Eq.~\eqref{eq:En-thermal}, the sum runs
over $e$ and the two heavy flavours, and the primes denote the mixed values.

\paragraph{Non-uniform conversion} Fast flavour conversions depend on position and
direction~\cite{Lund:2025jjo}. We let the conversion probability depend on where and how each
neutrino leaves the disc, $P=P_0\,p(r_{\rm em},\mu_{\rm em})$, with $\mu_{\rm em}$ the cosine between the
ray and the disc normal in the frame of the emitting gas; at the equator $p_\theta^2=\eta E^2$ and
the gas moves azimuthally, so that $\mu_{\rm em}=\sqrt\eta/(r_{\rm em}\,\varepsilon_{\rm em}/E)$. The map of
Eq.~\eqref{eq:mixing} is then applied packet by packet, which still conserves the number of
neutrinos. The engine stores the moments of each species weighted by $1$, $\rho$, $\mu_{\rm em}$ and
$\rho\mu_{\rm em}$, with $\rho=\exp[-(r_{\rm em}-r_{\rm isco})/r_{\rm isco}]$, so that any $p$ in their span
is evaluated exactly after the ray tracing. We use six profiles: uniform ($p=1$), concentrated in
the inner disc ($p=\rho$) or in the outer disc ($1-\rho$), along the normal ($\mu_{\rm em}$), at
grazing angles ($1-\mu_{\rm em}$), and at grazing angles in the inner disc ($\rho(1-\mu_{\rm em})$). This is a linear family of prescriptions, not an arbitrary dependence on position,
energy and direction.

\paragraph{An exact separability limit} If the heavy-lepton species are not emitted
($f_{\nu_x}=f_{\nub_x}=0$) and $P$, $\bar P$ are uniform in space and direction, Eq.~\eqref{eq:mixing} gives $f'_{\nu_e}=(1-P)f_{\nu_e}$ and
$f'_{\nu_x}=(P/2)f_{\nu_e}$, and every term of the rate is proportional to the same
$\mathcal M$ built from $\nu_e$ and $\nub_e$. Then
\begin{equation}\label{eq:ratio-nox}
\mathcal R=(1-P)(1-\bar P)+\frac{D_x}{2D_e}\,P\bar P
\end{equation}
exactly, whatever the geometry of the neutrinospheres and the spacetime. Equipartition
($P=\bar P=2/3$) gives $\mathcal R=0.159$ for $\sin^2\theta_W=0.23$, a reduction by a factor 6.3,
comparable to the reductions reported for black-hole discs~\cite{Uribe:2019cpq}. This holds even
when the $\nu_e$ and $\nub_e$ distributions differ in spectrum and angle, and then $\delta=0$
identically. The geometry cancels because every term of the rate involves the same pairs; if $P$
varied with the emission point or direction, the angular integrals would not cancel in general.
Equation~\eqref{eq:ratio-nox} provides an exact separability limit. The remaining question is how
rapidly this property breaks down once heavy-lepton emission becomes appreciable and its spatial
distribution differs from that of the electron flavours (Section~\ref{sec:flav-res}).

\section{Numerical method}\label{sec:numerics}

The calculation is split between a C library, parallelised with OpenMP over the points of the
map, and a Python layer that sets up the emitters, combines the moments into rates for any set
of species and spectra, integrates over volume and makes the figures. The code, ANNIHILUS~\cite{annihilus},
with the unit tests and the scripts that produce every table and figure of this paper, is publicly
available under the MIT licence.

\paragraph{Backward ray tracing} A direction on the local sky is given by
$\mu=\bOmega\cdot\mathbf e_r$ and an azimuth $\varphi$ around $\mathbf e_r$, so that
$\dd\Omega=\dd\mu\,\dd\varphi$. The ray lies in the plane of $\mathbf e_r$ and of the backward
direction. Starting from $u=1/r$ and $u'=\mathrm{sgn}(\mu)\sqrt{A}\,|\mu|/(r\sqrt{1-\mu^2})$, the
orbit equation is integrated with a fourth-order Runge--Kutta scheme in the orbital angle, with
step $h\min(1,u/|u'|)$ so that nearly radial rays keep the same relative accuracy ($h=0.01$).
Crossings of the equatorial plane occur at known orbital angles, $\varphi_k=\varphi_0+k\pi$; the
integrator steps exactly onto them and compares the crossing radius with the disc. For the sphere,
the fate of a ray that reaches its pericentre is decided exactly from the first integral
$u_p^2(1-2Mu_p)=b^{-2}$. Inside the photon sphere an ingoing ray is captured, and outside the
emitter and the photon sphere an outgoing ray escapes, which ends most rays after a few steps.

In Kerr spacetime we integrate the second-order forms of the Mino-time equations,
$r''=\tfrac12\,\dd R/\dd r$ and $x''=-(\eta+\lambda^2-a^2)x-2a^2x^3$, which pass through turning
points without special treatment. The step $h/(|r'|/r+\sqrt{|\eta|+\lambda^2})$ reduces to the
orbital-angle step of the Schwarzschild tracer. Equatorial crossings are found from the change of
sign of $x$ and located by Newton iterations on the length of the Runge--Kutta step. Rays end at
$r=r_+(1+10^{-6})$, or when they move outwards beyond both the disc and $r=4M$.

\paragraph{Edge-adapted quadrature} The integrand is smooth on the sky except across the edges
of the images of the emitter. For each azimuth we scan $\mu$ on a grid uniform in polar angle
(64 points), refined geometrically towards both poles, where the image of a distant emitter and
the shadow sit. Each ray is classified by its outcome, by the number of equatorial crossings and
by whether each crossing passed inside or outside the disc. Crossing radii vary continuously
with direction, so a disc image that lies between two samples of the same order changes the class
however thin it is; recursive bisection then locates the edges between classes to $10^{-12}$ in
$\mu$. An island of a different class lying entirely between two samples of the same class could
still be missed, so we verify the procedure empirically below. Each segment that ends on the
emitter is integrated with 16 Gauss--Legendre nodes, and the azimuths with the periodic
trapezoidal rule (64 points).

\paragraph{Cost} A point costs $10^4$--$5\times10^4$ rays, about 0.2--0.7\,s on one core, almost all
of it in the ray tracing: the moment contraction takes less than a millisecond, whereas the
direct double sum over $10^4$ rays takes 1.7\,s (\ref{app:bench}). The gain is in the
angular kernel, not in the tracer.

\paragraph{Volume integrals} We use Gauss--Legendre nodes in $\ln(r-2M)$ on radial panels whose
edges include $r_{\rm in}$ and $r_{\rm out}$, and in $\cos\theta$ on the upper hemisphere.

\paragraph{Tests} For the sphere the moments of ranks 0 to 3 agree with the analytic cone to
$10^{-9}$--$10^{-12}$, and $F(M/R)$ computed from the traced moments agrees with the analytic
integral to $10^{-12}$. In flat space the moments of an annulus agree with direct integration over
its area to $10^{-6}$--$10^{-10}$. Doubling the scan or quadrupling the azimuths changes $\dot q$
near the horizon only in the seventh digit. Rates for points mirrored across the equator agree
to $10^{-7}$. The Kerr tracer at $a=0$ reproduces the Schwarzschild tracer to $10^{-6}$--$10^{-9}$.
Kerr spacetime with $M=0$ is flat space in oblate spheroidal coordinates, so that every spin
term of the geodesic equations is active while the rays are straight; the moments of an annulus
then agree with direct integration to $2\times10^{-6}$ for $h=0.01$ and $10^{-7}$ for $h=0.005$,
with the $h^4$ convergence of the integrator. As an independent check of the Kerr potentials and of
the initial data, we integrate Hamilton's equations for null geodesics in Boyer--Lindquist
coordinates, with the inverse metric written out and differentiated numerically, without the
Carter constant or Mino time: for $a=0.9$ the radii of the first equatorial crossing agree with the
tracer to $10^{-9}$--$10^{-13}$, and captures and escapes coincide. Along the rays the first
integrals $(\dd r/\dd\sigma)^2=R(r)$ and $(\dd x/\dd\sigma)^2=\Theta_x(x)$ are preserved to
$3\times10^{-9}$. The projection $-Q_\mu\xi^\mu$ of Eq.~\eqref{eq:Einf-kerr}, built with the explicit
tetrad and metric, and $\sqrt{-g}=\Sigma\sin\theta$ are checked to $10^{-10}$.

\paragraph{Narrow images} Refining the scan in $\mu$ from 16 to 256 points changes the moments by
less than $10^{-8}$ at points near the axis, near the disc and near the hole, for $a=0$ and 0.9.
Images of order one and higher (rays that cross the equatorial plane outside the disc before
hitting it) carry 0.5--11\% of the moments at these points; their contribution from the quadrature
agrees with brute-force sampling by $2.6\times10^6$ rays on a uniform grid within the sampling
error of the latter (a few per cent of that contribution).

\section{Results}\label{sec:results}

\subsection{Validation}\label{sec:valid}

\begin{figure}[t]
\centering
\includegraphics{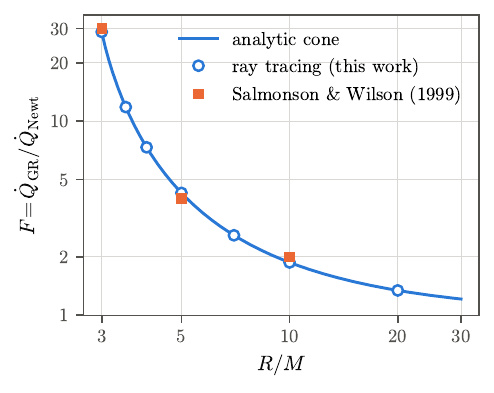}
\caption{Relativistic enhancement of the energy deposition around a static isothermal sphere at
fixed luminosity at infinity, integrated over proper volume. Line: analytic cone; circles: ray
tracing; squares: values quoted in Ref.~\cite{Salmonson:1999es}.}
\label{fig:sphere}
\end{figure}

\paragraph{Sphere} Figure~\ref{fig:sphere} shows the relativistic enhancement at fixed luminosity
at infinity. The traced values coincide with the analytic integral: $F=1.87$, 4.27, 7.33, 11.8 and
28.8 for $R/M=10$, 5, 4, 3.5 and 3, in agreement with the values quoted by Salmonson and
Wilson~\cite{Salmonson:1999es}.

\paragraph{Thin disc} Birkl et al.~\cite{Birkl:2006mu} considered an isothermal, non-rotating
ring between $6M$ and $7.7M$ around a black hole of $2\,M_\odot$, emitting $\nu_e$ and $\nub_e$ at
$T=5\times10^{10}$\,K (model D), the same ring in flat space (DN), and spheres of radius $3.4M$
with and without gravity (S and SN). Table~\ref{tab:birkl} compares our rates. With their
definitions, the agreement is within the few per cent expected from their Monte Carlo sampling;
the rates in proper volume depend on the inner boundary near the horizon, which we place at
$1.01\times2M$; the energy at infinity changes by less than 0.1\% when the boundary moves to
$1.0001\times2M$ and by 0.7\% when it moves out to $1.05\times2M$. With our rates, the exact energy at infinity, Eq.~\eqref{eq:Einf}, is 24\% lower than
$\dot E^{\infty}$ in their definition for the whole volume and 16\% lower for the region of
outgoing momentum. The relativistic gain over flat space for this ring is then 1.9 for the whole
volume and 1.6 for the outflow region, instead of 2.5 and 1.9.

\begin{table}[t]
\centering
\caption{Energy deposition rates ($10^{49}$\,erg\,s$^{-1}$) for the models of
Ref.~\cite{Birkl:2006mu}. ``Tot'' integrates over the whole volume and ``up'' over the region where
the radial momentum deposition is positive. The columns $\dot E^\infty$ use the definition of
Ref.~\cite{Birkl:2006mu}, $\int\sqrt A\,\dot q\,\dd V$; the last two columns give the energy at
infinity, Eq.~\eqref{eq:Einf}.}
\label{tab:birkl}
\begin{tabular}{llcccccc}
\toprule
model & & $\dot E^{\rm tot}$ & $\dot E^{\rm up}$ & $\dot E^{\rm tot,\infty}$ & $\dot E^{\rm up,\infty}$ & $\dot E_\infty^{\rm tot}$ & $\dot E_\infty^{\rm up}$\\
\midrule
D  & Ref.~\cite{Birkl:2006mu} & 0.54 & 0.27 & 0.36 & 0.23 & & \\
   & this work & 0.56 & 0.26 & 0.38 & 0.22 & 0.29 & 0.19\\
DN & Ref.~\cite{Birkl:2006mu} & 0.15 & 0.12 & 0.15 & 0.12 & & \\
   & this work & 0.15 & 0.12 & 0.15 & 0.12 & 0.15 & 0.12\\
S  & Ref.~\cite{Birkl:2006mu} & 0.12 & & & & & \\
   & this work & 0.12 & & & & & \\
SN & Ref.~\cite{Birkl:2006mu} & 0.066 & & & & & \\
   & this work & 0.066 & & & & & \\
\bottomrule
\end{tabular}
\end{table}

\paragraph{Spin} Ref.~\cite{Birkl:2006mu} also varied the spin for rings of zero angular
momentum (models DA1, REF, A.5 and A1; we use $a=0.9999$ for their $a=1$). Their local observers are
at rest in Boyer--Lindquist coordinates; such static observers exist only outside the ergosphere,
which Ref.~\cite{Birkl:2006mu} therefore excludes from the integrals. For this comparison we
transform our rates to the static frame, $\dot q_{\rm stat}=-Q_t/\sqrt{-g_{tt}}$, with $-Q_t$ as in
Eq.~\eqref{eq:Einf-kerr}; the radial components, and hence the region of outgoing momentum, are the
same as for the ZAMO, because the two frames differ by an azimuthal boost. We integrate over
$r>M+(M^2-a^2\cos^2\theta)^{1/2}$, with nodes in $(r-r_{\rm E})^{1/2}$ next to the static limit
$r_{\rm E}$, where $\dot q_{\rm stat}$ diverges as $(-g_{tt})^{-1/2}$ but remains integrable.
Table~\ref{tab:spin} shows that all four rates then agree with theirs to about 5\% (at most
5.1\%), as at $a=0$;
refining the volume grid ($14\times20$ instead of $10\times14$ nodes) changes the whole-volume
rates of DA1 and A.5 by at most 0.3\% and those of the outflow region, whose boundary only the grid
resolves, by up to 2.4\%. Our
own rates are those of ZAMOs, which extend into the ergosphere down to the horizon. Over the whole
volume, the rate in proper volume and the integral of $\sqrt{-g}\,\dot q$ are then dominated, for
$a\to1$, by the long throat outside the horizon, and depend on the inner boundary of the
integration: for A1 with the boundary at $1.01r_+$ they are 250 and 8 times the rates defined by
Ref.~\cite{Birkl:2006mu}, and moving the boundary from $1.001r_+$ to $1.0001r_+$ doubles the former, while the
exact energy at infinity changes by 0.5\%. This divergence belongs to these integrals extended into
the ergosphere, not to the procedure of Ref.~\cite{Birkl:2006mu}. The outflow region hardly feels
the spin: $\dot E^{\rm up}_\infty$ changes by less than 1\% from $a=0$ to $a\to1$, while the
whole-volume $\dot E_\infty$ grows by 27\% (REF to A1) and 38\% (D to DA1), from deposition close to
the horizon. This growth comes from the ergosphere: outside it, $\dot E_\infty$ exceeds its value
at $a=0$ by less than 1\% for both A1 and DA1.

\begin{table}[t]
\centering
\caption{Spin series of Ref.~\cite{Birkl:2006mu} ($10^{49}$\,erg\,s$^{-1}$): rings of zero angular
momentum at $T=5\times10^{10}$\,K, between $6M$ and $7.7M$ ($M=2\,M_\odot$, models D and DA1) and
between $6M$ and $10M$ ($M=3\,M_\odot$, REF, A.5, A1). The first four columns use the definitions of
Ref.~\cite{Birkl:2006mu}: rates measured by static observers, integrated outside the ergosphere
over proper volume ($\dot E^{\rm tot}$, $\dot E^{\rm up}$) and with the weight $\sqrt{-g}$
($\dot E^{\rm tot,\infty}$, $\dot E^{\rm up,\infty}$), over the whole volume (tot) or the region of
outgoing radial momentum (up). The last two columns give the exact energy at infinity,
Eq.~\eqref{eq:Einf-kerr}, down to the horizon, with the inner boundary at $1.0001r_+$ for $a=0.9999$
and $1.01r_+$ otherwise. Model D is recomputed here on the volume grid of the spin series; its rates
in the region of outgoing momentum differ from those of Table~\ref{tab:birkl} by about 2\%, within
the sensitivity of that region to the volume grid noted in the text.}
\label{tab:spin}
\begin{tabular}{lllcccccc}
\toprule
model & $a$ & & $\dot E^{\rm tot}$ & $\dot E^{\rm up}$ & $\dot E^{\rm tot,\infty}$ & $\dot E^{\rm up,\infty}$ & $\dot E_\infty^{\rm tot}$ & $\dot E_\infty^{\rm up}$\\
\midrule
D   & 0      & Ref.~\cite{Birkl:2006mu} & 0.54 & 0.27 & 0.36 & 0.23 & & \\
    &        & this work & 0.56 & 0.27 & 0.38 & 0.23 & 0.29 & 0.19\\
DA1 & 0.9999 & Ref.~\cite{Birkl:2006mu} & 0.56 & 0.27 & 0.38 & 0.23 & & \\
    &        & this work & 0.59 & 0.27 & 0.40 & 0.22 & 0.40 & 0.19\\
REF & 0      & Ref.~\cite{Birkl:2006mu} & 8.7  & 5.1  & 6.2  & 4.4  & & \\
    &        & this work & 8.9  & 5.3  & 6.4  & 4.6  & 5.1  & 3.9\\
A.5 & 0.5    & Ref.~\cite{Birkl:2006mu} & 8.9  & 5.1  & 6.3  & 4.4  & & \\
    &        & this work & 9.1  & 5.3  & 6.5  & 4.6  & 5.1  & 3.9\\
A1  & 0.9999 & Ref.~\cite{Birkl:2006mu} & 8.9  & 5.1  & 6.5  & 4.4  & & \\
    &        & this work & 9.3  & 5.3  & 6.7  & 4.5  & 6.5  & 3.9\\
\bottomrule
\end{tabular}
\end{table}

\subsection{How gravity redistributes the deposition}\label{sec:maps}

To separate gravity from the special-relativistic Doppler shifts of the flow, we compare each
relativistic calculation with the same discs in flat space ($M=0$, straight rays) moving with the
same local speeds $v(r)$ as the relativistic disc (Section~\ref{sec:flavour}), for which
$\mathsf g=\sqrt{1-v^2}/(1-v\lambda/r_{\rm em})$, and with static discs in flat space.

\begin{figure}[t]
\centering
\includegraphics{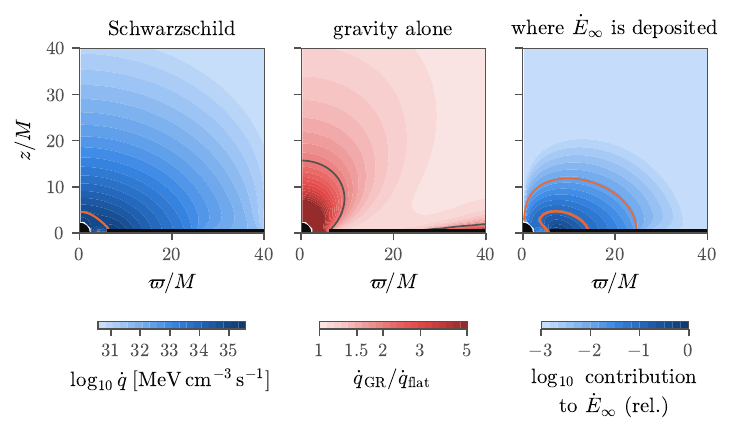}
\caption{Left: energy deposition rate in the static frame for $\nu_e\nub_e$ around a Keplerian
disc (thick line) in Schwarzschild spacetime; the orange line separates inward from outward radial
momentum deposition. Centre: ratio to the same disc in flat space, moving with the same local
velocities; the grey line marks a ratio of 2. Right: contribution to the energy at infinity per unit
area of the meridional plane, $\varpi\sqrt{A}\,\dot q$ with $\varpi=r\sin\theta$, relative to its maximum; the orange lines
enclose 50\% and 90\% of the total within the map. The rates depend on $M$ only through the length
unit.}
\label{fig:maps}
\end{figure}

Figure~\ref{fig:maps} shows $\dot q$ for $\nu_e\nub_e$ around a Keplerian disc in Schwarzschild
spacetime, with $r_{\rm in}=6M$, $\nu_e$ emitted out to $40M$ and $\nub_e$ out to $25M$, at 4 and
5\,MeV at the inner edge and $T\propto r^{-1/2}$, and its ratio to the same disc in flat space.
Gravity raises the rate everywhere: by 10--20\% far from the hole, by more than a factor 2 within
about $8M$ of it and up to $15M$ along the axis, by up to a factor 5 above the inner disc, and by factors 2--4
just above the outer disc, where light bending lifts the hot inner disc, seen at grazing incidence
in flat space, into view. Rotation alone acts differently: in flat space it lowers the rate on the
axis to 0.5--0.7 of the static value, through the transverse Doppler redshift of the inner disc,
and raises it by up to 20\% just above the inner disc, through beaming. Radial momentum is deposited inwards only in a small region next
to the hole (orange line). High local rates do not mean a large share of the total: weighting by
the volume element (right panel), half of $\dot E_\infty$ within the map comes from $r<10M$ and
90\% from $r<20M$, mostly from the layer within $5M$ of the disc (73\%); the region within $5M$ of
the hole, where gravity raises the rate most, contributes 6\%. Since the amplification varies so
strongly in space, it need not be common to neutrinos emitted at different radii, and hence to
different flavours.

\subsection{Flavour conversion}\label{sec:flav-res}

Whether gravity amplifies all flavour populations alike decides if flavour conversion can be
treated without it. We first examine
overlapping emitting regions, where a common geometric weighting is expected to be a useful
approximation, and then separate the heavy-lepton and electron-flavour neutrinospheres to test
its breakdown.

\begin{figure}[t]
\centering
\includegraphics{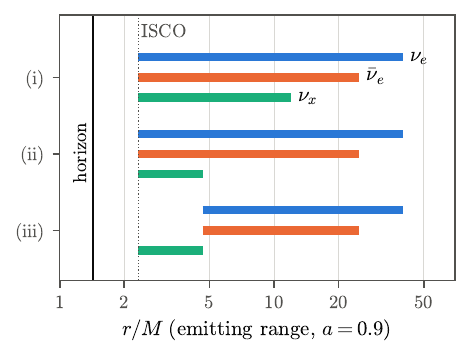}
\caption{Emitting ranges of $\nu_e$, $\nub_e$ and $\nu_x$ in the three layouts, shown for $a=0.9$.
In (i) all neutrinospheres overlap in the inner disc; in (ii) the heavy-lepton emission is
confined to an inner ring; in (iii) the ring and the electron-flavour discs do not overlap.}
\label{fig:layouts}
\end{figure}

\paragraph{Set-up} Each species has its own Keplerian disc, in three layouts (Fig.~\ref{fig:layouts}):
(i) all discs start at the ISCO and end at $40M$ ($\nu_e$), $25M$ ($\nub_e$) and $12M$ ($\nu_x$);
(ii) as (i), with $\nu_x$ emitted only within $2r_{\rm isco}$; (iii) as (ii), with $\nu_e$ and
$\nub_e$ emitted only beyond $2r_{\rm isco}$, so that the heavy-lepton and electron-flavour
neutrinospheres do not overlap. The reference temperatures are 4, 5 and 6\,MeV for $\nu_e$, $\nub_e$
and $\nu_x$ at the ISCO (at the flux peak for the Page--Thorne profile), with $T\propto r^{-1/2}$,
$r^{-1/4}$, $r^{-3/4}$ or $\mathcal F^{1/4}$. For each case we sweep 36 combinations of
$T_x/T_e\in\{0.8,1,1.25,1.5\}$, $T_{\bar e}/T_e\in\{1,1.25,1.5\}$ and a grey factor
$\epsilon_x\in\{0.1,0.3,1\}$ that scales the heavy-lepton emission (these only rescale the stored
moments), the six conversion profiles of Section~\ref{sec:flavour}, and the $36\times6$
combinations of the two. We consider $a=0$, 0.9 and 0.99, and compare with the same discs in flat
space, moving with the same local velocities, through $\delta$, Eq.~\eqref{eq:delta}, with flat
discs rotating at Newtonian Keplerian speeds, and with static discs.

\begin{figure}[t]
\centering
\includegraphics{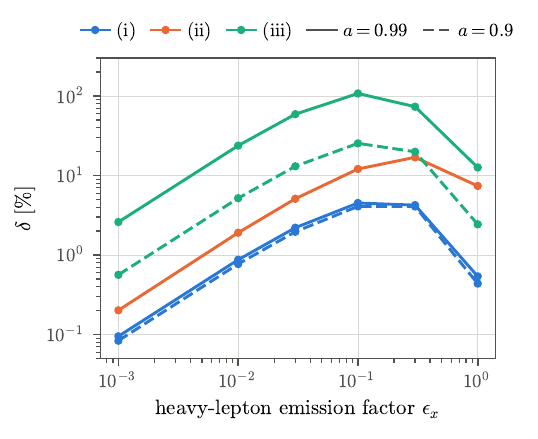}
\caption{Gravitational change of the conversion factor, $\delta$, as a function of the strength of
the heavy-lepton emission, $\epsilon_x$ ($\epsilon_x=0$ is the exact limit, $\delta=0$), at the reference
temperatures, for the three layouts at $a=0.99$ (solid) and for layouts (i) and (iii) at $a=0.9$
(dashed).}
\label{fig:emission}
\end{figure}

\paragraph{From the exact limit} Without heavy-lepton emission, $\delta$ vanishes by
Eq.~\eqref{eq:ratio-nox}, and the computed $\delta$ is zero to numerical precision. Figure~\ref{fig:emission}
shows how the separability breaks down as the heavy-lepton emission is switched on through
$\epsilon_x$, at the reference temperatures. Starting from the exact separability limit at vanishing heavy-lepton emission, $\delta$
initially increases with $\epsilon_x$, reaches a maximum at intermediate emission strengths,
$\epsilon_x\simeq0.1$--$0.3$, and subsequently decreases. The growth is slow for
overlapping neutrinospheres, where $\delta$ stays below 5\%, and fast for segregated ones: in layout
(iii) at $a=0.99$, $\delta=24\%$ for $\epsilon_x=0.01$ and 108\% for $\epsilon_x=0.1$. Weak but not negligible
heavy-lepton emission, as in black-hole discs, is thus where gravity affects the conversion factor
most.

\begin{figure}[t]
\centering
\includegraphics{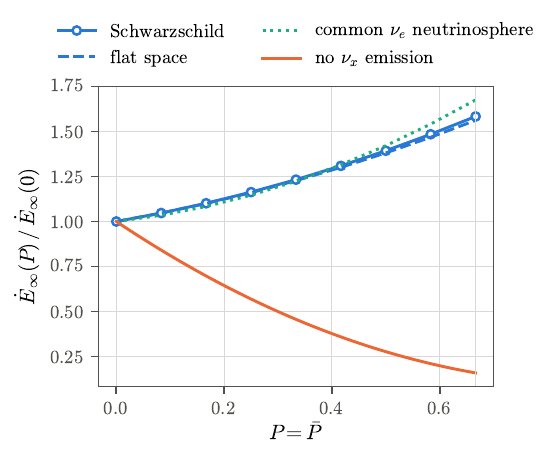}
\caption{Energy at infinity with flavour conversion relative to no conversion, for layout (i)
around a Schwarzschild black hole and for the same discs in flat space, moving with the same local
velocities. Dotted: all
species on the $\nu_e$ disc. Orange: no heavy-lepton emission, Eq.~\eqref{eq:ratio-nox}.}
\label{fig:flavour}
\end{figure}

Figure~\ref{fig:flavour} shows $\mathcal R(P)$ for layout (i) at $a=0$: because the
heavy-lepton species are hotter, conversion raises the rate (by 58\% at equipartition), whereas
without heavy-lepton emission it lowers it by a factor 6.3. The relativistic factor differs from
the flat-space one by at most 1.4\%; assuming instead that all species share the $\nu_e$ disc
changes it by 6\%.

\begin{figure}[t]
\centering
\includegraphics[width=\textwidth]{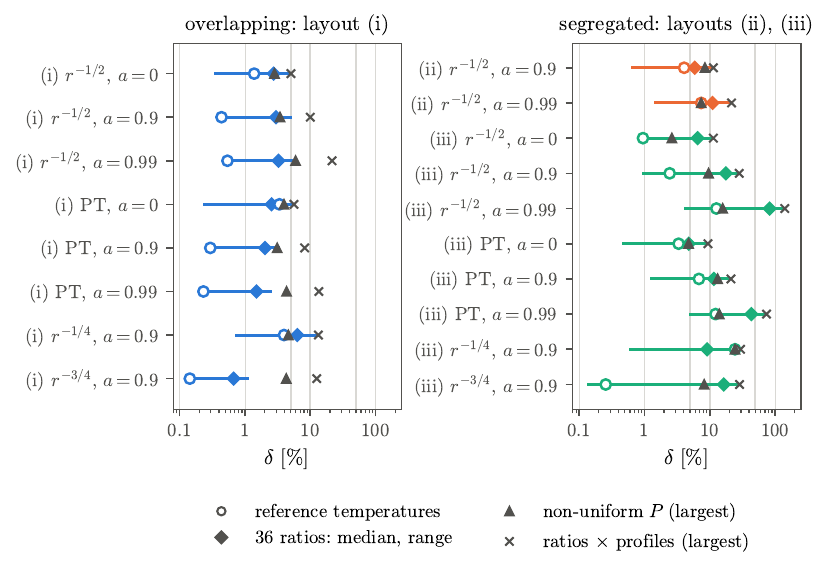}
\caption{Change of the flavour-conversion factor due to gravity, $\delta$, for overlapping (left)
and segregated (right) neutrinospheres, for the temperature profiles and spins studied: circles
for the reference temperatures, diamonds and lines for the median and range over 36 temperature
and luminosity ratios with uniform conversion, triangles for the largest value over the six
conversion profiles at the reference temperatures, crosses for the largest value over
the 36 ratios crossed with the six profiles. Flat-space discs with the same local velocities.
Vertical lines mark 5, 10 and 50\%.}
\label{fig:robust}
\end{figure}

\paragraph{Overlapping neutrinospheres} Table~\ref{tab:flavour} and Fig.~\ref{fig:robust} (left)
summarise layout (i). For uniform conversion, gravity changes the conversion factor by at most 6\%
over the 36 temperature and luminosity ratios and the three spins for the $r^{-1/2}$, $r^{-3/4}$
and Page--Thorne profiles, and by up to 13\% for the flattest profile, $r^{-1/4}$, in which the
outer discs, where the species differ, weigh more. Meanwhile it raises the deposition by factors
1.4--4.7. The six conversion profiles, at the reference temperatures, also stay within 6\%. The
two tests are separate. For the $r^{-1/2}$, $r^{-3/4}$ and Page--Thorne profiles, crossing the
conversion profiles with the 36 ratios leaves $\delta$ below 6\% at $a=0$ but raises it to
8--13\% at $a=0.9$ and 14--22\% at $a=0.99$, always for weak and cold heavy-lepton emission
($\epsilon_x=0.1$, $T_x=0.8\,T_e$); for $r^{-1/4}$ it stays at 13\%. There the rate is dominated by the electron flavours, which
are emitted over a range of radii that gravity amplifies differently, and conversion concentrated
in the inner or the outer disc, or in some directions, weights these radii differently; the exact
limit of Section~\ref{sec:flavour} requires uniform conversion. In this regime the flat-space
factor can thus be applied to a relativistic rate computed without conversion, to a few per cent,
for uniform conversion; how far the conversion may depart from uniformity with this accuracy is
not established by the six profiles, and would require a control of the amplitude of the
non-uniformity. Departures below about 0.3\% are at the level of the
numerical error of $\delta$ (\ref{app:bench}) and carry no physical meaning.

\paragraph{Kinematics of the flat-space disc} Two refinements of the comparison come before the
segregated layouts. Compared with static discs, the conversion factor changes by up to 20\%
(layout~(iii), $a=0.99$), partly through the special-relativistic Doppler shifts of the fast inner
disc: flat-space estimates must include the motion of the emitter. They must also use its local
speeds. The Newtonian Keplerian speeds exceed them near the ISCO of a rapidly spinning hole (0.83
instead of 0.60 for $a=0.99$), so that the inner disc is beamed more strongly in flat space, which
mimics part of the gravitational enhancement: in layout (iii) at $a=0.99$, $\delta$ at the
reference temperatures falls from 13\% to 2.5\%, and its largest value over the sweep from 141\%
to 83\% (last column of Table~\ref{tab:flavour}). For overlapping neutrinospheres and uniform
conversion the two flat-space references give $\delta$ within one percentage point of each other.
Non-uniform conversion at the reference temperatures gives at most 6\% for overlapping
neutrinospheres and 24\% for the segregated ones discussed next.

\begin{figure}[t]
\centering
\includegraphics[width=\textwidth]{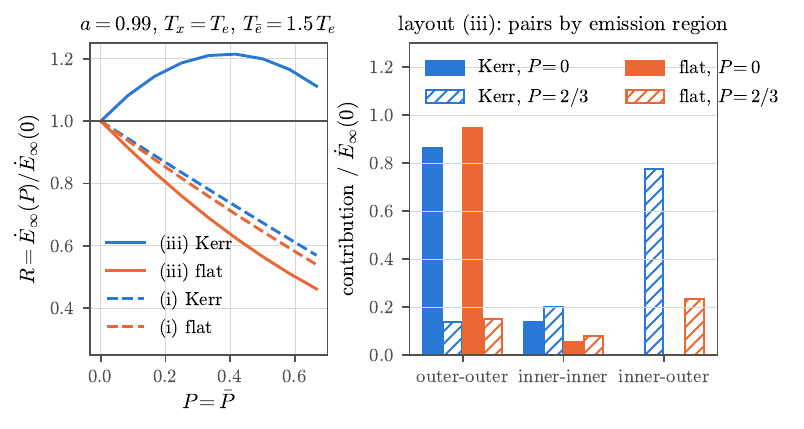}
\caption{Sign reversal for $a=0.99$, $T_x=T_e$, $T_{\bar e}=1.5T_e$. Left: conversion factor in Kerr
spacetime and for the same discs in flat space, moving with the same local velocities, for the
segregated layout (iii) (solid) and
the overlapping layout (i) (dashed). Right: deposition in layout (iii) split by the emission
regions of the two partners (outer disc, inner ring), in units of the unconverted deposition, for
$P=0$ (filled) and $P=2/3$ (hatched).}
\label{fig:reversal}
\end{figure}

\paragraph{Segregated neutrinospheres} When the heavy-lepton neutrinos come from an inner ring
inside the electron-flavour neutrinospheres (Fig.~\ref{fig:robust}, right), gravity changes the
conversion factor by up to 21\% in layout (ii) and 141\% in layout (iii) at $a=0.99$, where the
relativistic factor is 2.4 times the flat-space one, and by up to 74\% for the Page--Thorne
profile. Segregation alone does not fix $\delta$: at the reference temperatures and
$\epsilon_x=1$, layout (iii) with $T\propto r^{-1/2}$ gives only $\delta=0.9\%$ at $a=0$ and 2.4\% at
$a=0.9$ (13\% at $a=0.99$), whereas weaker heavy-lepton emission (Fig.~\ref{fig:emission}) or hotter $\nub_e$
raise it to tens of per cent. The spectra and relative intensities of the species, and not only
their layout, set $\delta$. To see why gravity matters so much here, we split the deposition of
layout (iii) into the three kinds of pairs it contains (Fig.~\ref{fig:reversal}, right): both
partners from the outer disc, both from the inner ring, or one from each. Gravity raises the first
by a factor 2.0, and the deposition by pairs involving inner-ring neutrinos by factors 5.7--7.4: neutrinos from the
ring reach the deposition region, which lies mostly within $10M$ of the hole and close to the disc
(Fig.~\ref{fig:maps}, right), along strongly bent paths and with large energy shifts. Without
conversion, only the outer-outer and inner-inner pairs contribute, and the former dominate.
Conversion turns heavy-lepton neutrinos from the ring into electron flavour, which then annihilate
with outer-disc $\nub_e$ (and $\nu_e$) through the larger coupling $D_e$: it moves the deposition
towards the pairs that gravity amplifies most.

\paragraph{Sign reversal} More importantly, the breakdown of separability is not merely a
quantitative correction: general relativity can reverse whether flavour conversion enhances or
suppresses neutrino annihilation. For $T_x=T_e$ and $T_{\bar e}=1.5T_e$ around a hole with $a=0.99$,
equipartition lowers the deposition by 54\% in flat space but raises it by 11\% in Kerr spacetime
(Fig.~\ref{fig:reversal}, left). The reversal follows from the cross pairs identified above.
Conversion depletes the outer-disc $\nu_e\nub_e$ pairs, which gravity amplifies by 2.0, and creates
pairs of inner-ring neutrinos with outer-disc antineutrinos, and the reverse, which it amplifies
by 6.3--7.4. At equipartition these inner-outer pairs carry 0.23 of the unconverted deposition in
flat space, too little to compensate for the loss, but 0.77 in Kerr spacetime, enough to turn the
balance (Fig.~\ref{fig:reversal}, right). The reversal does not depend on the flat-space
kinematics: with Newtonian Keplerian speeds the reduction in flat space is 39\%, and the cross
pairs, amplified by 3.8--4.6, carry 0.36 of the unconverted deposition. It is, however, a result
of the transport model, in which converted neutrinos are not reabsorbed
(Section~\ref{sec:flavour}); conversion on a surface outside the opaque regions should be tested
before it is read as a prediction for real discs.

\begin{table}[t]
\centering
\caption{Gravity and flavour conversion. Gain: deposited Killing energy relative to the flat-space
discs with the same local velocities and to static flat-space discs, without conversion. $\delta$,
Eq.~\eqref{eq:delta}: change of the conversion factor due to gravity (relativistic versus flat space
with the same local velocities) for the reference temperatures, for $\bar P=P$ and, in parentheses,
for $\bar P$ fixed by lepton-number conservation, $P(N_{\nu_e}-N_{\nu_x})=\bar P(N_{\nub_e}-N_{\nub_x})$,
over the values of $P$ with $\bar P\le1$ (in layout (i) with $a=0.99$, with PT at $a=0.9$, 0.99
and with $r^{-3/4}$ this restricts $P$ to below 0.08--0.4; ``--'': no $\bar P>0$ conserves the
lepton number, because $N_{\nu_e}-N_{\nu_x}$ and $N_{\nub_e}-N_{\nub_x}$ have opposite signs);
``total'': versus static discs, i.e.\ gravity and rotation; ``profiles'': largest over the six
conversion profiles; ``sweep'': median and largest over 36 temperature and luminosity ratios, with
uniform conversion; ``$\times$prof.'': largest over the 36 ratios crossed with the six profiles;
``Kepl.'': largest over the 36 ratios with Newtonian Keplerian speeds in flat space; in the
columns ``static'' and ``total'', ``--'' means not computed. PT:
Page--Thorne profile. At $a=0$ layout (ii) coincides with (i), because $2r_{\rm isco}=12M$, the
outer edge of the $\nu_x$ disc in (i).}
\label{tab:flavour}
\scriptsize\setlength{\tabcolsep}{2.5pt}
\begin{tabular}{llcccccccccc}
\toprule
 & & & \multicolumn{2}{c}{gain} & \multicolumn{7}{c}{$\delta$ [\%]}\\
\cmidrule(lr){4-5}\cmidrule(lr){6-12}
layout & $T(r)$ & $a$ & flat & static & reference & total & profiles & sweep med. & sweep max & $\times$prof. & Kepl.\\
\midrule
(i) & $r^{-1/2}$ & 0 & 1.56 & 1.57 & 1.4 (0.6) & 1.2 & 2.8 & 2.7 & 5.1 & 5.1 & 4.6\\
(i) & $r^{-1/2}$ & 0.9 & 2.21 & 2.46 & 0.4 (--) & 0.4 & 3.4 & 3.0 & 4.9 & 10 & 4.8\\
(i) & $r^{-1/2}$ & 0.99 & 4.72 & 5.29 & 0.5 (1.3) & 0.3 & 6.0 & 3.3 & 5.5 & 22 & 4.6\\
(i) & PT & 0 & 1.42 & -- & 3.4 (3.0) & -- & 4.0 & 2.6 & 5.6 & 5.6 & 5.6\\
(i) & PT & 0.9 & 1.99 & -- & 0.3 (0.5) & -- & 3.1 & 2.0 & 3.3 & 8.2 & 3.2\\
(i) & PT & 0.99 & 3.54 & -- & 0.2 (0.2) & -- & 4.3 & 1.5 & 2.5 & 14 & 2.7\\
(i) & $r^{-1/4}$ & 0.9 & 1.81 & -- & 4.0 (3.5) & -- & 4.6 & 6.3 & 13 & 13 & 12\\
(i) & $r^{-3/4}$ & 0.9 & 2.44 & -- & 0.1 (0.1) & -- & 4.3 & 0.7 & 1.1 & 13 & 1.1\\
(ii) & $r^{-1/2}$ & 0.9 & 2.33 & 2.44 & 4.0 (2.8) & 4.0 & 8.4 & 5.9 & 11 & 11 & 11\\
(ii) & $r^{-1/2}$ & 0.99 & 5.10 & 5.27 & 7.4 (6.5) & 6.8 & 7.4 & 11 & 21 & 22 & 17\\
(iii) & $r^{-1/2}$ & 0 & 1.58 & 1.56 & 0.9 (1.0) & 1.8 & 2.6 & 6.5 & 11 & 11 & 14\\
(iii) & $r^{-1/2}$ & 0.9 & 2.38 & 2.38 & 2.4 (1.2) & 8.8 & 9.6 & 18 & 28 & 28 & 25\\
(iii) & $r^{-1/2}$ & 0.99 & 5.03 & 4.94 & 13 (13) & 20 & 16 & 83 & 141 & 141 & 83\\
(iii) & PT & 0 & 1.42 & -- & 3.3 (3.0) & -- & 4.7 & 4.8 & 9.4 & 9.4 & 11\\
(iii) & PT & 0.9 & 2.07 & -- & 6.8 (5.9) & -- & 13 & 12 & 21 & 21 & 19\\
(iii) & PT & 0.99 & 3.73 & -- & 12 (14) & -- & 14 & 43 & 74 & 74 & 48\\
(iii) & $r^{-1/4}$ & 0.9 & 1.69 & -- & 24 (24) & -- & 24 & 9.1 & 30 & 30 & 27\\
(iii) & $r^{-3/4}$ & 0.9 & 2.53 & -- & 0.3 (--) & -- & 8.2 & 16 & 28 & 29 & 24\\
\bottomrule
\end{tabular}
\end{table}

\FloatBarrier
\section{Conclusions}\label{sec:conclusions}
We have identified the spatial arrangement of flavour-dependent emitting regions as a key
determinant, in the models considered, of whether flavour-conversion corrections to
neutrino pair annihilation above black-hole discs can be computed without gravity. The tool
behind this result combines backward ray tracing in Schwarzschild and Kerr spacetimes with an
exact angular-moment evaluation of the annihilation kernel; it reproduces the analytic sphere and
the disc models of Ref.~\cite{Birkl:2006mu}, including their spin series, gives the deposited
Killing energy exactly, and makes the angular integration a negligible part of the cost. The
results follow a single progression.

(1) \emph{Exact limit.} For spatially and angularly uniform conversion and negligible
heavy-lepton emission, the conversion factor does not depend on the spacetime or on the emission
geometry, Eq.~\eqref{eq:ratio-nox}.

(2) \emph{Overlapping emitters.} When the neutrinospheres of all flavours extend to the inner
disc and the conversion is uniform, separability holds approximately: over the temperature
profiles and the temperature and luminosity ratios explored, gravity changes the conversion factor
by at most 6\% (13\% for a very flat temperature law), while it changes the deposition itself by
factors 1.4--5 for spins up to 0.99. Conversion that varies with the emission radius or direction,
together with weak heavy-lepton emission, raises the change to 8--13\% at $a=0.9$ and up to 22\% at
$a=0.99$. The flat-space factor must include the Doppler shifts of the disc flow, with its local
speeds; they change it by up to 20\%.

(3) \emph{Segregated emitters.} When the heavy-lepton neutrinos come from an inner ring inside
the electron-flavour neutrinospheres, the conversion factor can change by tens of per cent, up to
a factor 2.4 around a hole with $a=0.99$ (1.8 if the flat-space disc is given Newtonian Keplerian
speeds), and conversion can turn from suppressing to enhancing the deposition. The change is
largest for weak but non-negligible heavy-lepton emission, and it depends on the spectra and
intensities of the species as well as on their layout.

(4) \emph{Mechanism.} Gravity amplifies the annihilation of pairs involving inner-ring neutrinos
about three times as much as that of outer-disc pairs, and conversion moves the deposition from the
latter to the former, through the larger electron-flavour coupling.

(5) \emph{Domain of application.} Flat-space conversion factors can be used in relativistic
estimates when the emitting regions overlap and the conversion is uniform; when they are
segregated, the converted species should be transported relativistically, and the radial layout of
the neutrinospheres in radiation-hydrodynamics simulations of post-merger discs indicates which
case is closer. These conclusions hold for thin discs, free streaming without reabsorption of
the converted neutrinos and conversion parametrised at emission, and they refer to the
deposited Killing energy, which is an energy budget rather than the energy of an observed outflow;
the conversion is phenomenological and does not replace collective flavour evolution. A conversion
surface above the disc, neutrinospheres and spectra taken from simulations, and
neutrino-dominated accretion flows computed with neutrino microphysics are the next tests.

\section*{Acknowledgements}
This study was financed in part by the Coordena\c{c}\~ao de Aperfei\c{c}oamento de Pessoal de
N\'ivel Superior -- Brasil (CAPES) -- Finance Code 001. L.O.J. and P.M.B. acknowledge financial
support from the Funda\c{c}\~ao de Amparo \`a Pesquisa e Inova\c{c}\~ao do Estado de Santa Catarina
(FAPESC). R.C.R.L. acknowledges support from PROPPG No.~15/2025 -- PROINT-UDESC and thanks
SciCom-UFES for the computational resources.

\appendix

\section{Energy at infinity}\label{app:einf}

Let $T^{\mu\nu}$ be the stress--energy tensor of the pair plasma and
$Q^\mu\equiv\nabla_\nu T^{\mu\nu}$ the four-momentum given to it per unit proper four-volume by
annihilation. For an observer with four-velocity $u^\mu$ and orthonormal triad $e_{\hat\imath}^\mu$,
$Q^\mu=\dot q\,u^\mu+c\dot p^{\hat\imath}e_{\hat\imath}^\mu$, where $\dot q$ and $c\dot p^{\hat\imath}$ are
Eqs.~\eqref{eq:factorised} and \eqref{eq:pdot} evaluated with the distributions measured by that
observer: the integrand of Eq.~\eqref{eq:qdot-kinetic}, with $\varepsilon+\varepsilon'$ replaced by
$p^\mu+p'^\mu$, is a four-vector, because the bracket and $\dd^3p/\varepsilon$ are invariant.

Let $\xi=\partial_t$ be the timelike Killing vector, normalised to $\xi\cdot\xi=-1$ at infinity.
Since $\nabla_{(\mu}\xi_{\nu)}=0$, $\nabla_\nu(T^{\mu\nu}\xi_\mu)=Q^\mu\xi_\mu$. Integrating over the
region between the slices $t$ and $t+\dd t$ outside the horizon and using Gauss's theorem,
\begin{equation}
\frac{\dd}{\dd t}\int_{\Sigma_t}(-T^t{}_\mu\xi^\mu)\sqrt{-g}\,\dd^3x+\Phi_\partial
=\int_{\Sigma_t}(-Q_\mu\xi^\mu)\sqrt{-g}\,\dd^3x ,
\end{equation}
where $\Phi_\partial$ is the flux of Killing energy through the horizon and the outer boundary;
annihilation supplies Killing energy to the plasma, per unit time at infinity, at the rate
$\dot E_\infty=\int(-Q_t)\sqrt{-g}\,\dd^3x$. In Boyer--Lindquist coordinates $\sqrt{-g}=\Sigma\sin\theta$
and, for the ZAMO frame, $\partial_t=\alpha\,e_{\hat t}-\omega\sqrt{g_{\phi\phi}}\,e_{\hat\phi}$, hence
\begin{equation}\label{eq:Einf-kerr}
\dot E_\infty=\int(\alpha\,\dot q+\omega\sqrt{g_{\phi\phi}}\;c\dot p_{\hat\phi})\,\Sigma\sin\theta\,\dd r\,\dd\theta\,\dd\phi ,
\end{equation}
which is Eq.~(16) of Ref.~\cite{Zalamea:2010ax}. In Schwarzschild spacetime $\alpha=\sqrt A$, $\omega=0$
and $\Sigma\sin\theta=r^2\sin\theta$, which gives Eq.~\eqref{eq:Einf}. No assumption on the momentum
of the pairs enters. The integrand of Ref.~\cite{Birkl:2006mu}, $\sqrt{-g}\,\dot q$, lacks the lapse
$\alpha$ and the frame-dragging term; Ref.~\cite{Birkl:2006mu} evaluates it with the rate of static
observers and outside the ergosphere (Section~\ref{sec:valid}). Evaluated instead with the ZAMO rate
and extended into the ergosphere of a nearly extremal hole ($a=0.9999$, model A1), its volume
integral grows by a factor 11 when the inner boundary moves from $1.1r_+$ to $1.0001r_+$, while
Eq.~\eqref{eq:Einf-kerr} changes by 15\% and converges (0.5\% between $1.001r_+$ and $1.0001r_+$).

\section{Cost and convergence}\label{app:bench}

\paragraph{Moment contraction versus double sum} At a point above a Keplerian disc around a Kerr
hole ($a=0.9$, $r=6M$, $\theta=0.8$) we trace $N$ rays on a Fibonacci sphere, with equal weights,
and evaluate $\dot q$ both by the direct double sum over pairs of rays and by the moment contraction
of Eq.~\eqref{eq:factorised}, on one core. Table~\ref{tab:bench} lists the times. The two
evaluations agree to $10^{-15}$, as they must. The double sum grows as $N^2$ and exceeds the ray
tracing for $N\gtrsim10^3$; the contraction stays below a millisecond up to $N=10^4$ and is a few
per cent of the tracing time at $N=10^5$. The equal-weight sampling converges slowly because of the
edges of the image: with $N=10^5$ rays it agrees to $2\times10^{-4}$ with the edge-adapted
quadrature of Section~\ref{sec:numerics}, which uses $2\times10^4$ rays and 0.7\,s.

\begin{table}[h]
\centering
\caption{Time per point (ms, one core) for $N$ rays on a Fibonacci sphere: ray tracing, direct
double sum and moment contraction, and the error of the equal-weight sampling relative to the
edge-adapted quadrature. The double sum was not computed for $N=10^5$.}
\label{tab:bench}
\begin{tabular}{rcccc}
\toprule
$N$ & tracing & double sum & moments & sampling error\\
\midrule
$10^2$ & 1.3 & 0.13 & 0.17 & $5\times10^{-2}$\\
$10^3$ & 12 & 30 & 0.17 & $2\times10^{-2}$\\
$10^4$ & 121 & 1700 & 0.41 & $3.5\times10^{-3}$\\
$10^5$ & 1200 & -- & 74 & $2\times10^{-4}$\\
\bottomrule
\end{tabular}
\end{table}

\paragraph{Convergence of the global rates} For the Keplerian disc of Section~\ref{sec:results}
around a hole with $a=0.9$ (inner edge at the ISCO, outer edge at $40M$, $T\propto r^{-1/2}$), we
change one numerical parameter at a time from the values used in this paper. $\dot E_\infty$ changes
by $10^{-9}$ when the scan in $\mu$ is halved or doubled, by $2\times10^{-5}$--$4\times10^{-5}$ when the
radial or polar nodes of the volume integral change by 30--50\%, by $3\times10^{-5}$ when the
integration step is doubled, by $3\times10^{-4}$ when the Gauss--Legendre nodes per segment double
and by $10^{-3}$ when the azimuths double. The relativistic rates are thus accurate to about
$10^{-3}$, limited by the azimuthal rule, whose error decreases roughly as the inverse square of the number of azimuths. The rate
in the region of outgoing momentum is less accurate, about 1\%, because the boundary of that region
is resolved only by the volume grid. The flat-space rates converge more slowly: in the refinement
below, the rates without conversion change by at most $1.5\times10^{-3}$ in curved spacetime, but
by $2$--$4.5\times10^{-3}$ in flat space with the local speeds, and by up to 1.2\% with Newtonian
Keplerian speeds around $a=0.99$. The conversion factors are ratios of rates computed with the same
quadrature, so part of these errors cancels; the next paragraph measures what remains.

\paragraph{Convergence of $\delta$} We recomputed the conversion factors with all quadratures
refined at once: twice the azimuths (64 to 128) and the Gauss--Legendre nodes per segment (16 to
32), and a finer volume grid (14 instead of 10 radial nodes per segment, 24 instead of 16 polar
nodes). We chose the configurations of Table~\ref{tab:flavour} with the smallest $\delta$
(layout~(i), $T\propto r^{-3/4}$, $a=0.9$), an intermediate one (layout~(ii), $a=0.9$) and the
largest (layout~(iii), $a=0.99$), each in curved spacetime and in both flat-space cases. Over the
36 combinations of temperature and luminosity ratios, $\delta$ changes by at most 0.13, 0.36 and
0.28 percentage points in the three configurations; the largest value goes from 140.8\% to
140.5\%. For the non-uniform conversion profiles the changes stay below 0.5 percentage points. In
the sign-reversal case of Fig.~\ref{fig:reversal}, the change of the deposition at equipartition
stays at $+11.2\%$ in Kerr and at $-53.8\%$ in flat space ($-53.7\%$ refined). With Newtonian
Keplerian speeds in flat space the changes are larger, up to 1.0 percentage point over the sweep
(82.5\% to 81.5\%) and 1.5 for the profiles, and the change at equipartition goes from $-39.1\%$
to $-38.7\%$. A $\delta$ of a few per cent thus carries a relative error of up to $\sim20\%$, but
of at most half a percentage point (one with Newtonian Keplerian speeds), so the separation between
the overlapping and segregated regimes and the sign reversal are converged. Two resolutions measure
the stability of $\delta$, not a strict bound on the remaining error. Finally, the maximum in Eq.~\eqref{eq:delta} is taken over nine equally spaced values
of $P$. For $\bar P=P$ the rates are quadratic polynomials in $P$, so the maximum over the
continuous interval follows exactly from three of them; it exceeds the nine-point value by less than
0.05 percentage points in all configurations, ratios and profiles, for both flat-space references,
and $\mathcal R_{\rm flat}\ge0.16$ throughout. For $\bar P$ fixed by lepton-number conservation,
where the range of $P$ can end below $2/3$, at $P_{\rm max}=1/k$ with $k=\bar P/P$, the rates are
also quadratic in $P$, and the value in Table~\ref{tab:flavour} is the exact maximum over
$0\le P\le P_{\rm max}$: the largest of the values at the two ends and at the stationary points of
the ratio of the two quadratics.

\bibliographystyle{elsarticle-num-jheap}
\bibliography{references}

\end{document}